\documentclass[useAMS,usenatbib]{mn2e} 
\usepackage{aas_macros}
\usepackage{graphics}
\usepackage[pdftex]{graphicx}
\usepackage{epstopdf}
\usepackage{epsfig}  
\usepackage{natbib} 
\usepackage{float}
\usepackage{amsmath}
\usepackage{times}
\usepackage[varg]{txfonts}
\usepackage{verbatim} 
\usepackage{multirow,bigdelim} 
\usepackage{color}
\usepackage[normalem]{ulem}

\def  \LCDM{$\Lambda$CDM}

\def \kms {{\rm km s$^{-1}$}}

\newcommand{\hmpc}{{\,\rm h^{-1}Mpc}}
\def\br{{\bf r}}

\def\bv{{\bf v}}

\title[Goodness-of-fit analysis   of CF2]{Goodness-of-fit analysis  of the Cosmicflows-2 database of velocities}

\author[Hoffman et al.]
{Yehuda Hoffman$^1$, Adi Nusser$^2$
H\'el\`ene M. Courtois$^{3}$ and R. Brent Tully$^4$\\
$^1$Racah Institute of Physics, Hebrew University, Jerusalem 91904, Israel\\
$2$ Physics Department,  The Technion, Haifa 32000, Israel\\
$^3$University of Lyon; UCB Lyon 1/CNRS/IN2P3; IPN Lyon, France\\
$^4$Institute for Astronomy (IFA), University of Hawaii, 2680 Woodlawn Drive, HI 96822, USA\\
}

\begin{document}
\date{Apr. 2016}


\maketitle

\label{firstpage}

\begin{abstract}

The goodness-of-fit  (GoF) of the Cosmicflows-2 (CF2) database of peculiar velocities with the \LCDM\ standard   model  of cosmology  is presented. 
Standard application of the $\chi^2$ statistics of the full database, of its 4,838 data points, is hampered by the small scale non-linear dynamics which is not accounted for by the (linear regime) velocity power spectrum. The bulk velocity constitutes a highly compressed representation of the data which filters out the small scales non-linear modes. Hence the statistics of the bulk flow provides an efficient tool for assessing the GoF of the data given a model.
The particular approach  introduced  here is to use the (spherical top-hat window) bulk velocity extracted  from the Wiener filter reconstruction of the 3D velocity field 
as a linear low pass filtered highly compressed representation of the   CF2 data. An ensemble   2250 random linear realizations of the WMAP/\LCDM\    model  has been used to calculate the bulk velocity auto-covariance matrix.
We find   that the CF2 data is consistent with the WMAP/\LCDM\ model to better than the 2$\sigma$ confidence limits. This provides a further validation that the CF2 database is consistent with the standard model of cosmology.

\end{abstract}


\section{Introduction}
\label{sec:intro}

In the standard model of cosmology the large scale structure (LSS) of the universe grows out of a primordial perturbation field via gravitational instability. 
The continuity equation implies that the evolving density field is associated with a peculiar velocity field, both of which represent departures,  or fluctuations, from a pure Hubble expansion. The Cosmic Microwave Background (CMB) dipole anisotropy,  interpreted as the consequence of the peculiar motion of the Local group (LG) relative  to the CMB frame of reference is the best evidence and example of that motion. First hints for the dipole anisotropy were given by \cite{1969Natur.222..971C} and  \cite{1971Natur.231..516H} and a more definitive determination by \cite{1977PhRvL..39..898S}. In the standard model the density and the velocity fields are  connected and one can opt for a full description of the  LSS  of the universe by means of either the density or the velocity (or both) field \citep{1980lssu.book.....P}. Velocities of galaxies are taken here as the prime tracers of the LSS.

The quest for a mapping of the local universe has been a major driver of the study of the velocities of galaxies
\citep{1981ApJ...246..680T,
1982ApJ...258...64A,
1988ApJ...326...19L,
1990ApJ...364..370B,
1990ApJ...364..349D,
1995ApJ...454...15S,
1995MNRAS.276.1391N,
1995PhR...261..271S,
1997ApJS..109..333W,
2007ApJS..172..599S,
 2008ApJ...676..184T,
 2014MNRAS.445.2677S}.
 The study of the bulk velocity dipole has been part of a major  effort to do cosmology by velocities. The notion of a bulk velocity is that of the (weighted) mean velocity of a finite chunk of space, e.g. the mean velocity of a sphere of radius R centered on the LG. It follows that a bulk velocity is a moment taken over  the full velocity field. Alternatively, a bulk velocity can be viewed as a highly compressed representation of a    large database of velocities - namely a database of hundreds or thousands entries is compressed into three numbers, corresponding to the three Cartesian components of the bulk velocity vector. 
Indeed, velocity-based cosmology has been  heavily dominated by attempts to estimate the bulk velocity on various scales
\citep{2009MNRAS.392..743W,
2010ApJ...709..483L,
2010MNRAS.407.2328F,
2011MNRAS.414..264C,
2011ApJ...736...93N,
2011ApJ...735...77N,
2012MNRAS.420..447T,
2014MNRAS.437.1996M,
2015MNRAS.447..132W}. 
\citep[See][for a thorough analysis of different methods of estimation of the bulk velocity.]{2016MNRAS.455..178N}
The motivation for  studies of the bulk velocity is twofold. One is the wish to understand the origin of the CMB dipole and its relation and possibly coherence with the flow  field on larger scales. The other is to use it as a constraint on   cosmological models and the values of various cosmological parameters   
\citep{1995ApJ...455...26J,1988MNRAS.231..149K,
2009MNRAS.392..743W,
2011ApJ...736...93N,
2012MNRAS.420..447T,
2012MNRAS.425.1709M}.

In the standard model   structure has emerged from a primordial random Gaussian perturbations field. This implied that on scales larger than $\approx10 \hmpc$
 (where $h$ is Hubble's constant in units of $100$\kms/Mpc) the velocity field constitutes a random Gaussian vector field. This enables a powerful and rigorous approach to the problem of the reconstruction of the LSS - both the density and velocity fields - from a given database of peculiar velocities. 
 Given a database of peculiar velocities and assuming  a prior cosmological model  which  postulates that 
the underlying primordial perturbation field is Gaussian of a given  power spectrum,  the LSS is readily reconstructed by means of the Wiener filter (WF)  and constrained simulations (CRs) that sample the scatter around the mean (WF) field \citep{1991ApJ...380L...5H,1995ApJ...449..446Z,2001misk.conf..223H}. 
This WF/CRs Bayesian framework provides an appealing  framework for the reconstruction of the LSS from velocities data within the realm of the standard model of cosmology 
 \citep{1999ApJ...520..413Z,2001MNRAS.326..375Z,2002MNRAS.336.1234Z,2012ApJ...744...43C,2014Natur.513...71T,2015ApJ...812...17P}. The WF/CRs reconstruction of the 3D velocity field was used to estimate the local bulk flow on scale of up to a few hundreds of Mpc
\citep{2015MNRAS.449.4494H}.

The Bayesian approach is indeed robust and optimal, within the context mentioned above. It focuses on the reconstruction of the LSS within the framework of the standard model of cosmology and  for a given database.  However,  most previous studies have not addressed the question  how consistent is the  assumed cosmological model with the observed data. This is not a trivial issue - the model needs to agree with the data so as to provide a solid foundation for the WF/CRs construction. Ideally one should have started with  establishing the agreement of the model with the data and only then reconstruct the LSS in the manner described above. However, history does not always proceeds in a linear fashion. The aim of the paper is to amend that situation and establish  the likelihood of peculiar velocities databases   given the    standard model of cosmology. 
The relevant methodology is  straightforward and well established. One needs to calculate the likelihood function of the data given the model - namely the probability of the occurrence of the data within the framework of the assumed model  \citep{1995ApJ...449..446Z,1995ApJ...455...26J,2001misk.conf..223H,Press:2007:NRE:1403886}. 
The likelihood  function establishes the goodness-of-fit (GoF) of the data by the model. In the cosmological case and under the assumption of the linear regime, where the velocity field constitutes a random Gaussian vector field and the observational errors are normally distributed, the likelihood analysis amounts to calculating a $\chi^2$ statistics.   
This approach was indeed applied to velocity databases \citep{1997ApJ...486...21Z,2001MNRAS.326..375Z,1995ApJ...455...26J}. The application of the  likelihood analysis 
to actual velocity databases suffers however from one major drawback. The gravitational dynamics of structure formation induces  non-linear contributions to the velocities of galaxies.  These non-linear corrections render the parameter estimation and GoF analysis  to be rather uncertain. The remedy to the problem involves the filtering of small scales to give linearized data. The likelihood analysis can then be safely applied to the linearized data. Here we suggest such a small scales filtering procedure and study the extent by  which  Cosmicflows-2 (CF2)  \citep{2013AJ....146...86T} is compatible with the \LCDM\ standard model of cosmology. 

The paper is structured as follows:  The   Bayesian and likelihood approach is derived in \S  \ref{sec:stat}.
The issue of a possible circularity of the use of the WF to assess the GoF of the data by the model is discussed and refuted in \ref{sec:circ}.
The CF2 data is reviewed in \S  \ref{sec:cf2}.
and its application to the CF2 dataset is presented in \S \ref{sec:results}. The paper concludes with a discussion at \S \ref{sec:results}.

\label{sec:disc}

\section{Statistical analysis }
\label{sec:stat}

\subsection{Bayesian approach: The Wiener Filter}
\label{subsec:wf}

The problem addressed here is that of the estimation of the underlying 3D velocity field in general, and the bulk flow in particular, given the data. The Bayesian framework provides the means to do that (Zaroubi et al 1995).

Modelling the data:
\begin{equation}
\label{eq:data}
U_\mu=u_\mu + \epsilon_\mu  = \bv(\br_\mu)\cdot \hat{r}_\mu + \epsilon_\mu
\end{equation} 
Here $U_\mu$ is the $\mu$-th data point, corresponding to  the radial velocity component at $\br_\mu$ ($u_\mu$), and $\epsilon_\mu$ is the statistical observational error of that data point. (Here, $\mu, \nu = 1, ..., N_{\rm data}$ where $N_{\rm data}$ is the number of data points.)

The Wiener Filter (WF) estimator  of the 3D velocity field on a cartesian grid:
\begin{equation}
\label{eq:wf}
v{^{\rm WF}_\alpha}(\br_i) = \xi{^\alpha_\mu}(\br_i) \xi{^{-1}_{\mu\nu}} U_\nu
\end{equation}
Here $\br_i$ is the coordinate of the i-th grid point, $ \xi{^\alpha_\mu}(\br_i)$ is the cross-correlation function,
\begin{equation}
\label{eq:cross-corr}
\xi{^\alpha_\mu}(\br_i)   = \Big<  v_\alpha(\br_i) U_\mu    \Big> = \Big<  v_\alpha(\br_i) \bv(\br_\mu)\cdot \hat{r}_\mu    \Big>,
\end{equation}
and the data-data auto-covariance matrix is:
\begin{equation}
\label{eq:auto-cov}
 \xi_{\mu\nu}  =  \Big<  U_\mu U_\nu \Big>  = \Big<   u_\mu u_\nu\Big> + \big(\sigma{^2_i} + \sigma{^2_\ast}\big) \delta{^K_{\mu\nu}}
\end{equation}

Given the WF reconstruction of the 3D flow field a Bayesian estimation of the bulk flow is easily obtained. The bulk flow is defined here as the mean  velocity, in the sense of a top-hat window weighting, of a sphere of radius $R$, centred on the Local Group:
\begin{equation}
\label{eq:b}
B_\alpha = {1\over V_R} \int_{r < R} \bf{v}_\alpha(\br) {\rm d}^3 r
\end{equation}
(Here $V_R= (4 \pi / 3) R^3$.) In the case of the WF reconstruction and for a series of spheres of radii $R_a$ ($a=1, ..., N_{\rm spheres}$) one finds:

\begin{equation}
\label{eq:bwf}
B{^{WF}_{a,\alpha} } = {1\over V_{R_a}  }   \int_{r < R_a}  v{^{\rm WF}_\alpha} (\br)  \   {\rm d}^3 r  = 
{1\over V_{R_a}} \int_{r < R_a}  \xi{^\alpha_\mu}(\br) \  {\rm d}^3 r  \  \  \  \xi{^{-1}_{\mu\nu}} U_\nu 
\end{equation}

It should be noted that $B{^{WF}_{a,\alpha}} $ is a random Gaussian variable, drawn from a random Gaussian field. It is a vector Gaussian variable obtained by convolving a set of Gaussian variable, $\big\{ U_\mu\big\}$, with a  given linear operator. This linear operator is determined by the assumed prior model, the distribution of the data points and the errors model.

\subsection{ Likelihood analysis}
\label{subsec:like}

The problem addressed here is that of the degree of compatibility of the data with a given model, namely what is the likelihood of a given dataset to emerge from a given model.

A straightforward  approach is to apply  the likelihood analysis to the raw data \citep{1995ApJ...449..446Z}:
\begin{equation}
\label{eq:liker}
\mathcal{L}(\big\{ U_\mu\big\} \vert {\rm model}) = {1\over \sqrt{{\rm det}\big((\xi_{\mu\nu}) \big)}} \exp\big[ - {U_\mu \xi{^{-1}_{\mu\nu}} U_\nu \over 2}\big]
\end{equation} 

For Gaussian variables, such as $\big\{ U_\mu\big\} $,  the value of the $\chi^2$ and the number of the degrees of freedom, determine the likelihood of the data given the model. In the present case $\chi^2 = U_\mu \xi{^{-1}_{\mu\nu}} U_\nu $, the number of degrees of freedom is the number of data points and the model is embodied by the data auto-covariance matrix, $\xi_{\mu\nu}$. This approach was applied to various databases  \citep{1997ApJ...486...21Z,2001MNRAS.326..375Z,2002MNRAS.336.1234Z}. However, it turns out that this statistical analysis is sensitive to $\sigma_\ast$,  the small scale power induced by non-linear dynamics on small scales \citep{1995ApJ...455...26J,1997ApJ...486...21Z}. This sensitivity to   $\sigma_\ast$ renders the likelihood analysis of the raw velocity data quite futile.

A different approach to the likelihood estimation is to apply it to a linear low pass filtered data, thereby  removing or suppressing the small scale component that  introduces non-linearities to the data. The bulk flow estimator is an ideal  low pass filter - $B{^{WF}_{a,\alpha}} $ is affected predominantly by waves longer then  $  \approx R_a$.   It follows that a likelihood analysis that  is shielded from small scales non-linear dynamics is provided by,
  \begin{equation}
\label{eq:likeb}
\mathcal{L}(\big\{B{^{WF}_{a,\alpha}} \big\} \vert {\rm model}) = {1\over \sqrt{ {\rm det}\big( \Xi{^{WF}_{a,\alpha, b,\beta}}\big)   }} \exp\big[ - {B{^{WF}_{a,\alpha}} \   \Xi{^{WF}_{a,\alpha, b,\beta}}^{-1} \ B{^{WF}_{b,\beta}}  \over 2}\big],
\end{equation} 
where the bulk velocity auto-covariance matrix ($\Xi{^{WF}_{a,\alpha, b,\beta}}$) is
\begin{equation}
\label{eq:bbcov}
\Xi{^{WF}_{a,\alpha, b,\beta}}= \Big<B{^{WF}_{a,\alpha}} B{^{WF}_{b,\beta}} \Big> = {1\over V_{R_a} V_{R_b}}  \int_{r < R_a}\int_{r' < R_b}
\xi{^\alpha_\mu}(\br) \xi{^{-1}_{\mu\mu{\prime}}}  \xi{^\beta_{\mu^\prime}}(\br') \ {\rm d}^3 r   {\rm d}^3 r'.
\end{equation}

\section{Is the methodology circular?}
\label{sec:circ}

In the field of cosmology and LSS one often encounters the claim that the WF reconstruction in general, and of the bulk flow in particular cannot be used to assess cosmological parameters and GoF of peculiar velocity data by cosmological data. The scope and implications of the present section exceed and reach beyond the specific aim of the paper but it should be considered as one of our key  results.

The methodology of \citet{1995ApJ...449..446Z} is the basis for extracting information from  large scale structure observational data, where the underlying fields are assumed to be Gaussian. This methodology assumes a cosmological prior in terms of a power spectrum which allows an inference of 
velocity and density fields. The WF estimation of the bulk flow has appeared  in several other, sometime implicit,  forms. This is the case of the `all space constrained estimate' (ASCE) of \citet{2011ApJ...736...93N} and the minimal variance (MV) estimator of \citet{2015MNRAS.447..132W}. In fact, \citet{2016MNRAS.455..178N} has  demonstrated explicitly the tight relationship  of the  \citet{2015MNRAS.447..132W}  MV estimation with the WF, with its explicit dependence of an assumed power spectrum.

The WF is the optimal linear (in the data) estimator of the underlying field in the sense of the minimal variance between the   inferred  and true underlying field \citep{1992ApJ...398..169R,1995ApJ...449..446Z}. To the extent that the power spectrum and the errors are correctly known then no other linear estimator can perform better than the WF. Yet, even if a wrong model is assumed the resulting WF provides an estimator, but of a lesser accuracy. The following simple example makes the point.

Let us pose the following question. How does a wrongly chosen prior model for the WF affects the $\chi^2$  of the reconstructed field with respect to the assumed (wrong) model?

We illustrate the point with the following straightforward example. 
Consider $N$ data points $d_i=s_i+\epsilon_i$ where the random error $\epsilon_i$ is  normally distributed with variance $\sigma_\epsilon^2$. The underlying signal
$s_i$ 
obeys a Gaussian distribution of zero  mean and variance $\sigma_s^2$ and with no correlation between different points. Namely, both the signal and noise constitute a white noise.
Take a prior which assumes that the signal has a (wrong) mean $\bar s\ne 0$ and a variance $\sigma_s^2$, i.e. the correct value. 
Let $f_i$ be the WF estimated signal for point $i$, obtained by maximizing
\begin{equation}
2\ln P(f|d)\propto -(d_i-f_i)^2/\sigma_\epsilon^2-(f_i-\bar s)^2/\sigma_s^2     
\end{equation}
which gives
\begin{equation}
\label{eq:fi}
f_i=\frac{d_i \sigma{_s^2}+\bar{s} \sigma{_\epsilon^2}}   {\sigma{_s^2}+\sigma{_\epsilon^2}} \; .
\end{equation}
Let us explore the conditions under which the $f_i$ estimate is actually consistent with the assumed
prior. Following the methodology presented in the current paper we first compute the variance $\sigma{_f^2}=<f_i^2>$ {\it assuming }
the  prior, i.e. the average is taken as an  ensemble average  with  $d_i$ having   mean $\bar s$ and variance $\sigma_s^2$. It is easily seen that
\begin{equation}
\sigma_f^2=\sigma_s^2\left(\frac{\sigma_s^2}{\sigma_s^2 +\sigma_e^2}\right)^2 \; .
\end{equation}
Note that, as expected \citep[see][]{1995ApJ...449..446Z}, the variance of the WF estimate  of $s_i$ is biased.  
The $\chi^2$ of the estimated $f_i$ with respect to its expected variance is
\begin{equation}
\chi^2=\sum_i \frac{f_i^2}{\sigma_f^2},
\end{equation}
where $f_i$ is given by equation (\ref{eq:fi}) with $d_i$ being the actual data which has a mean of zero (in the ensemble average sense) and not  $\bar s$.
We get for the ensemble average of $\chi^2$ the expression
\begin{equation}
\chi^2=N(1+\frac{{\bar s}^2}{\sigma_s^2}) \; .
\end{equation}  
Thus, $\chi^2$ for the prior can significantly be different from the mean value $\chi^2=N$ depending on the ratio
$\bar s/\sigma_s$. 
It is trivial to show that if the correct model has been assumed as a prior the $\chi^2$ per degree of freedom would have been unity. It follows that the WF does not yield by construction a $\chi^2=N$. This is obtained only for the correct model. 

A further comment is due here.
If the data is of good quality then the WF recovery of the underlying signal
depends very weakly on the assumed prior. 
However in the cosmological case of peculiar velocities where   the signal to noise ratio deteriorates strongly with the distance the opposite occurs. In fact, it has been shown that a significant enhancement of the large scale power in the assumed prior hardly affects  the  estimated bulk flow  \citep{2011ApJ...736...93N}.

\begin{figure*}
\label{fig1}
\includegraphics[width=0.5\textwidth,angle=-90]{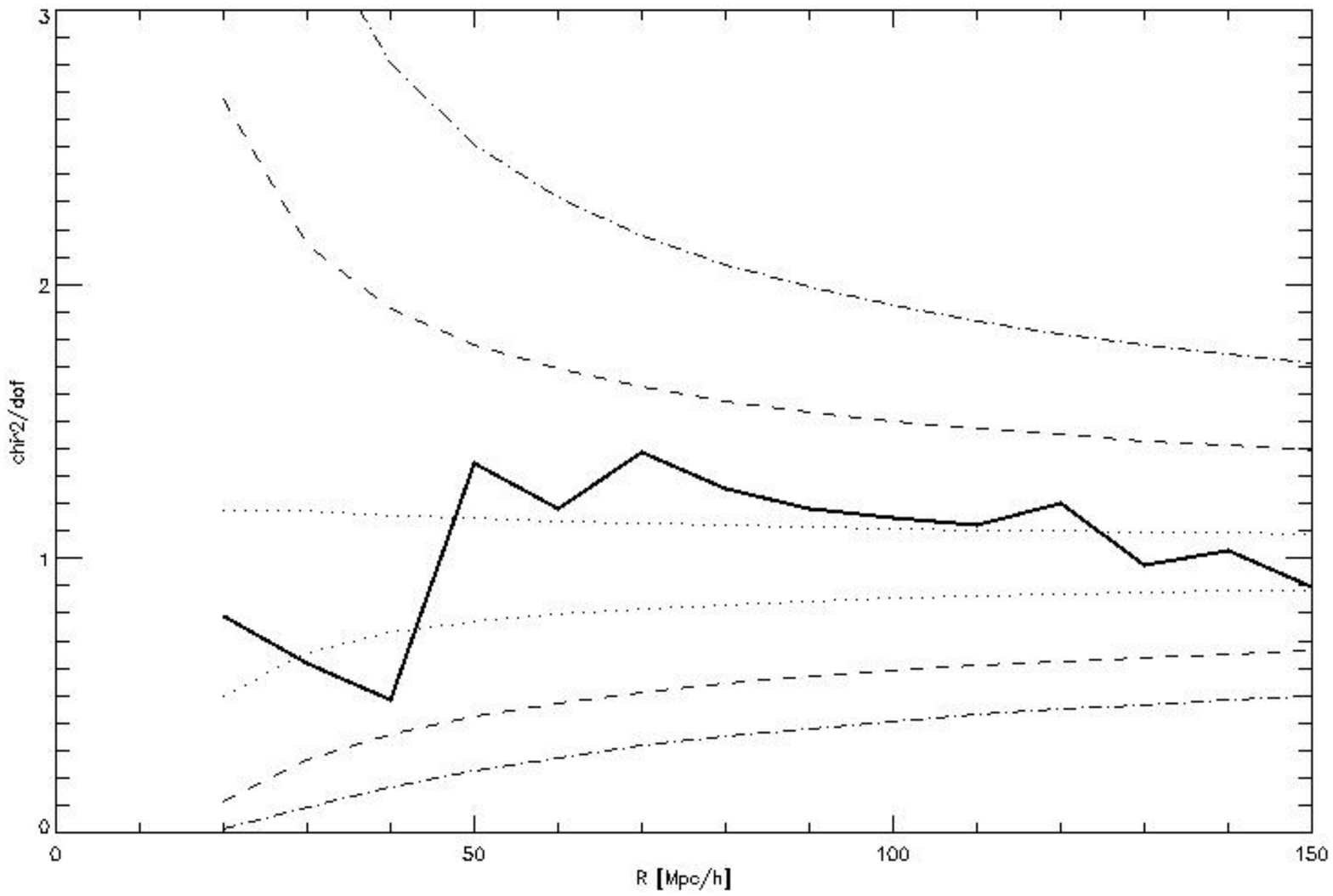}\\
\vspace{-0.5in}
\includegraphics[width=0.5\textwidth,angle=-90]{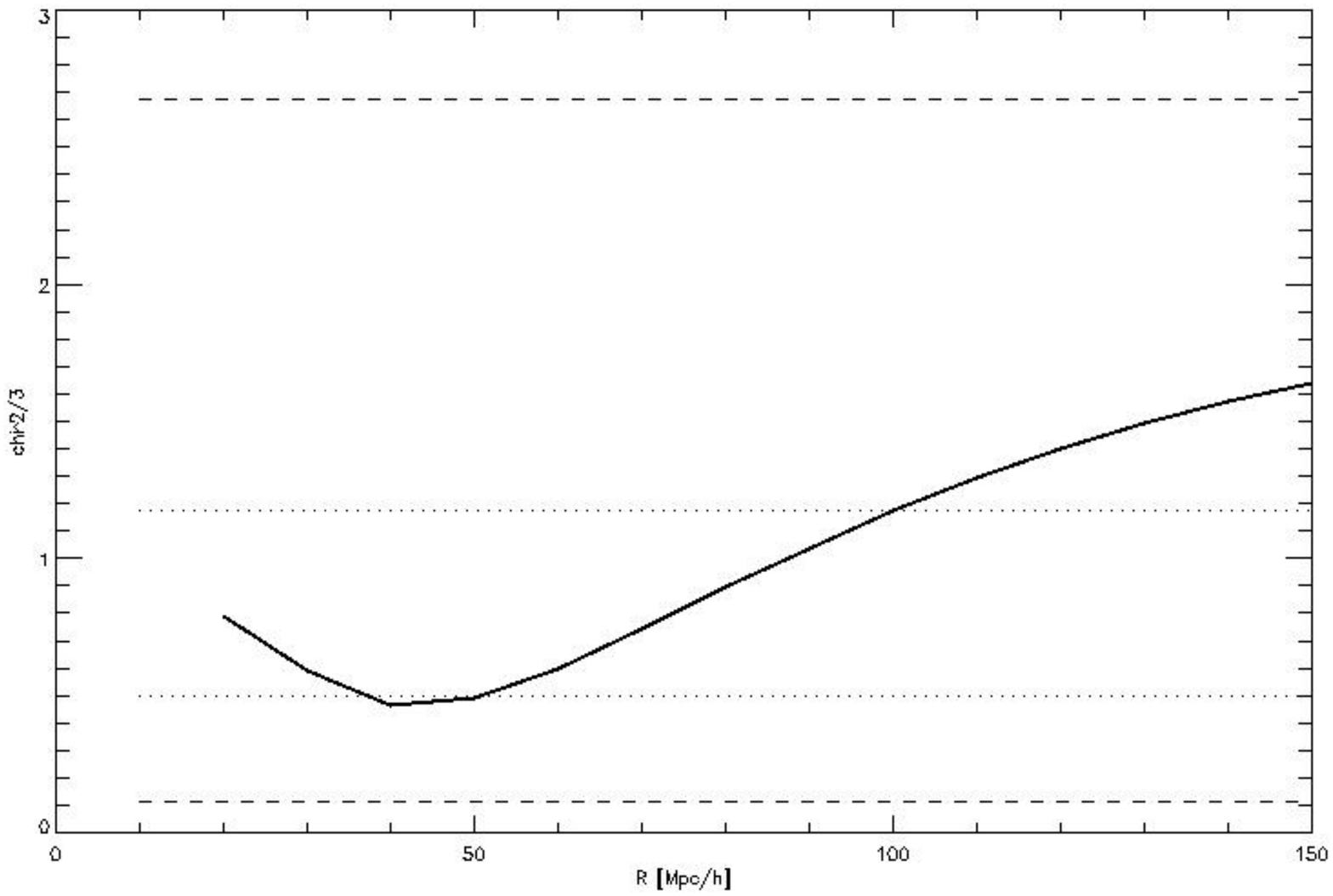}
\caption{ The $\chi^2$/d.o.f. of the bulk velocity vector of  concentric  spheres of radii $R=20, 30,$ ... $150\hmpc$. The bulk flow is calculated by convolving the Wiener filter velocity field with a spherical top-hat window function of radius $R$. The model, with respect to which the $\chi^2$ is evaluated, is the WMAP/\LCDM\ standard model of cosmology.  The confidence limits corresponding to the standard 1, 2 and 3$\sigma$ normal distribution are shown as well, namely the upper dotted, dashed and dot-dashed curves indicate the reduced $\chi^2$  for which the probability of not exceeding these values is 0.683, 0.945 and 0.997. The upper panel shows the cumulative  $\chi^2$ of all spheres of radii smaller or equal $R$, where the number of d.o.f. is 3  times the number of spheres. The lower panel shows the $\chi^2$ per 3 d.o.f. for individual spheres of radius $R$.
}
\end{figure*}

\section{Data}
\label{sec:cf2}

Cosmicflows-2 (CF2) is the second generation catalog of galaxy distances and velocities
built by the Cosmicflows collaboration \citep{2013AJ....146...86T}. The CF2 database  contains more than 8,000 galaxy peculiar velocities.
Distance measurements come mostly from the Tully-Fisher relation
and the Fundamental Plane methods \citep{2001MNRAS.321..277C}  Cepheids \citep{2001ApJ...553...47F}, Tip of the Red Giant
Branch \citep{1993ApJ...417..553L}, Surface Brightness Fluctuation \citep{2001ApJ...546..681T}, supernovae of type Ia \citep{2007ApJ...659..122J} and other miscellaneous methods also contribute to this large dataset but to a minor extent (12\%). 

A partial remedy to the problem of non-linearities which affect the linear WF/CRs reconstruction is provided by the  use of a grouped version of the  database  which contains 4,838 peculiar velocity values. The grouping is done by collapsing the velocities of all members of a given group or cluster of galaxies into one data point. The grouping is associated with the reduction of estimated error by the square root of the number of group members. The mean statistical error for individual galaxies of $\sim$ 20\% (of the distance) is thus reduced to a mean error for the groups of $\sim$ 9\%.

\section{Results}
\label{sec:results}

The standard model of cosmology, a flat $\Lambda$ cold dark matter (\LCDM) cosmology, is taken here as the prior model. The Wilkinson Microwave
Anisotropy Probe 5 (WMAP5) cosmological parameters are assumed \citep{2009ApJS..180..330K}. 
The following parameters, in particular, have been used here:  
$\Omega_m=0.28$ (the mass density parameter), $h=0.70$ (the Hubble constant in units of $100$\kms Mpc$^{-1}$) and $\sigma_8=0.817$ (the r.m.s. of the linear density fluctuations in a sphere of $8 \hmpc$). 
The model further assumes that the primordial fluctuations constitute a Gaussian random field.

The  bulk flow derived by  \citet{2015MNRAS.449.4494H} from the WF reconstruction of  the CF2 database  is used here as the low pass representation of the CF2 data.

The bulk velocity auto-covariance matrix ($\Xi{^{WF}_{a,\alpha, b,\beta}}$) has been evaluated as an ensemble average taken over 2250 random realizations of the linear velocity field. These realizations have been evaluated on a Cartesian 256$^3$ grid in a computational box of $L=1280\hmpc$. Each realization has been sampled at the position of the CF2 data points. Normal errors have been added according to the nominal errors of the CF2 individual data points. Each realization is yielding a mock of CF2 data - much in the same way as mock data is calculated  in the construction of CRs  \citep{1991ApJ...380L...5H}. The same WF operator that is applied to the actual CF2 database is applied to the mock catalog, yielding thereby  an ensemble of mock WF bulk flows evaluated for spheres of $R=20, 30,$ ... $150\hmpc$.   

Our intention here is to calculate the GoF of  the (WF bulk velocity of)  CF2 database with the WMAP/\LCDM\ model. It follows that we do not change the model and therefore  the pre-factor of the exponential term in the expression of the likelihood function (Eq. \ref{eq:likeb}) is a constant (for a given set of data) and is irrelevant. The quantity of interest is the reduced $\chi^2$, namely  $\chi^2/{\rm d.o.f.}$ where the number of degrees of freedom (d.o.f.) is the number of spheres considered times 3 (for the three cartesian components). Two variants are considered here. One is the cumulative $\chi^2$ of all sphere of radius smaller or equal R,
\begin{equation}
\label{eq:chi2c}
\chi^2(R_a,R_b \leq R) =   B{^{WF}_{a,\alpha}} \   \Xi{^{WF}_{a,\alpha, b,\beta}}^{-1} \ B{^{WF}_{b,\beta}},
\end{equation} 
and the $\chi^2$ of a given sphere of radius R,
\begin{equation}
\label{eq:chi2i}
\chi^2(R_a,R_b = R) =   B{^{WF}_{a,\alpha}} \   \Xi{^{WF}_{a,\alpha, b,\beta}}^{-1} \ B{^{WF}_{b,\beta}},
\end{equation} 

The upper panel of figure \ref{fig1} shows the cumulative $\chi^2$ for $R=20, 30,$ ... $150\hmpc$. The confidence limits corresponding to the standard 1, 2 and 3$\sigma$ normal distribution are shown as well, namely the upper dotted, dashed and dot-dashed curves indicate the reduced $\chi^2$  for which the probability of not exceeding these values is 0.683, 0.945 and 0.997 (values corresponding  to 1, 2 and 3$\sigma$ of the normal distribution). The allowed deviation from the expected reduced $\chi^2$ of unity decreases with radius as the number of d.o.f. increases.  The plot shows very clearly that the CF2 data (manifested by the WF bulk velocity) is confined  very comfortably within the 2$\sigma$ envelope. The $\chi^2/{\rm d.o.f.}$ of the individual spheres is shown here as well for the sake of reference (figure \ref{fig1}, lower panel). The confidence level now assume   constant (i.e. R independent) values as the number of d.o.f. is now fixed at 3 for all radii. Again the CF2 data  lies comfortably within the 2$\sigma$ confidence level.

\section{Discussion}
\label{sec:disc}
We show here a new approach to the linearization of peculiar velocity data and apply it to  the  GoF analysis of the database with a cosmological model. The motivation here is to bypass the bias introduced by the small scale non-linear dynamics which affects the more traditional likelihood, or $\chi^2$, analysis \citep[cf.][]{1995ApJ...449..446Z,1997ApJ...486...21Z,2001misk.conf..223H}. This is achieved by filtering the data of the small scale non-linear contributions. The particular approach that has been adopted here is to use the (spherical top-hat window) bulk velocity extracted  from the Wiener filter reconstruction of the 3D velocity field.  Our aim has been to assess by means of the $\chi^2$ statistics the compatibility of the Cosmicflows-2 (CF2) database  \citep{2013AJ....146...86T} with the WMAP/\LCDM\ cosmological model. It is found here that the CF2 data is consistent with the WMAP/\LCDM\ model to better than the 2$\sigma$ confidence limits. This provides a further validation that the CF2 database is consistent with the standard model of cosmology and can be used for the reconstruction of the LSS and the setting up of constrained  initial conditions for simulations that closely mimic the local universe. 

The methodology presented here can be  extended to perform a parameter estimation by performing a maximum likelihood analysis, using the likelihood function defined in Equation \ref{eq:likeb}.  Limiting ourselves to the realm of GoF analysis of the data given a model it should be noted that the use of the WF bulk velocity estimator   serves here primarily as a tool for the filtering of the small scales modes, and consequently the linearization of the data. The WF could have been evaluated with another cosmological model, not necessarily the \LCDM\ one and the results would have not changed by much. The use of the WF with the \LCDM\  model assures that the resulting estimated bulk velocity is the closest to the actual one - recalling that the WF is the minimal variance estimator - and thereby yielding the tightest possible GoF analysis among all    bulk velocity estimators.

Bulk velocities are a linear representation of a velocity database, typically made of thousands data points. As such they play a double role. They provide an efficient filtering of the small scales nonlinear modes. At the same time they constitute a drastic data compression and reduction the number of the d.o.f.  -   in the CF2 case the number goes from close to 5000 down to   a few tens. The current algorithm can be easily extended to include other moments of the velocity field - in particular the monopole and quadrupole terms. This will be provide a stricter GoF test on the data given a model. 

\bigskip
\begin{center}
\textbf{Acknowledgements}
\end{center}
AN has been  supported by the I-CORE Program of the Planning and Budgeting Committee and the Israel Science Foundation ( grants No. 1829/12 and No. 203/09).
YH has been partially supported by the Israel Science Foundation (1013/12). HC acknowledges support by the Institut Universitaire de France

\bibliography{biblicomplete}
\end{document}